\documentclass[acmsmall]{acmart}

\setcopyright{cc}
\setcctype{by}
\acmDOI{10.1145/3832211}
\acmYear{2026}
\acmJournal{PACMSE}
\acmVolume{3}
\acmNumber{ISSTA}
\acmArticle{ISSTA120}
\acmMonth{10}
\acmSubmissionID{issta26main-p1044-p}
\received{2026-01-27}
\received[accepted]{2026-06-25}

\usepackage{xcolor}
\usepackage{listings}
\usepackage{fancyvrb}
\usepackage{dirtree}
\usepackage{algpseudocode}
\usepackage{algorithm}
\usepackage{enumitem}
\usepackage{multirow}
\usepackage{tabularx}
\usepackage{booktabs}
\usepackage{pifont}
\setlist[itemize,1]{left=0pt}

\definecolor{darkgreen}{rgb}{0.0, 0.5, 0.0}

\lstdefinestyle{python}{
  language=Python,
  basicstyle=\ttfamily\small,
  keywordstyle=\color{blue}\bfseries,
  stringstyle=\color{red},
  commentstyle=\color{darkgreen},
  showstringspaces=false,
  breaklines=true,
  frame=single,
  numbers=left,
  numberstyle=\tiny\color{gray},
  stepnumber=1,
  numbersep=5pt,
  tabsize=2,
  captionpos=b,
  escapeinside={(*@}{@*)}
}

\newcommand{\code}[1]{{\ttfamily \small #1}}
\newcommand{\scode}[1]{{\ttfamily \footnotesize #1}}

\newcommand{\methodName}{IterTestQ}
\newcommand{\coreComponentName}{ITE process}

\newcommand{\TotalConfirmedOrFixedBugs}{17}

\newcommand{\TotalAllBugs}{23}
\newcommand{\TotalChangingConfiguration}{5}

\newcommand{\IdBugMissingStandardRxxGateDefinition}{\#2}
\newcommand{\IdBugUnrecognizedDelayGate}{\#3}

\newcommand{\IdBugMissingControlledPhaseCsGateDefinition}{\#5}

\newcommand{\IdBugQasmRegisterReorderingCausesSemanticChanges}{\#9}
\newcommand{\IdBugUnsupportedIswapGate}{\#10}

\newcommand{\IdBugIncorrectMcxGateGenerationCxInsteadOfCx}{\#13}

\newcommand{\IdBugDuplicateClassicalRegisterDefinition}{\#16}

\newcommand{\PercentageTimeITEprocess}{85.82\%}
\newcommand{\PercentageTimeBugDetection}{13.77\%}
\newcommand{\PercentageTimeGenerator}{0.41\%}
\newcommand{\PercentageTimetransform}{88.19\%}
\newcommand{\PercentageTimeexport}{6.27\%}
\newcommand{\PercentageTimeimport}{4.66\%}

\newcommand{\AvgGeneratorTimeOverall}{0.00089}
\newcommand{\TotalProgramsGeneratorOverall}{39,990}
\newcommand{\TotalTimeGeneratorOverall}{35.61}

\usepackage[utf8]{inputenc}
\usepackage[most]{tcolorbox}
\usepackage{xparse}
\usepackage{listings}
\usepackage{quantikz}

\usepackage{tikz}
\usetikzlibrary{fit}
\usetikzlibrary{shapes,arrows.meta,positioning}
\usetikzlibrary{backgrounds}

\definecolor{darkgray}{rgb}{0.4, 0.4, 0.4}
\lstdefinestyle{coolpython}{
    language=Python,
    basicstyle=\ttfamily\footnotesize,
    numbers=left,
    numberstyle=\ttfamily\footnotesize\color{darkgray},
    frame=single,
    xleftmargin=17pt,
    breaklines=true,
    showstringspaces=false,
    showtabs=false,
    tabsize=2,
    belowcaptionskip=5pt,
    backgroundcolor=\color{yellow!10},
    commentstyle=\color{orange!87!black}
}

\lstdefinestyle{coolqasm}{
    language=Python,
    basicstyle=\ttfamily\footnotesize,
    numbers=left,
    numberstyle=\ttfamily\footnotesize\color{darkgray},
    frame=single,
    xleftmargin=17pt,
    breaklines=true,
    showstringspaces=false,
    showtabs=false,
    tabsize=2,
    belowcaptionskip=5pt,
    backgroundcolor=\color{gray!10}
}

\lstdefinestyle{qasm}{
    language=Python,
    basicstyle=\ttfamily\footnotesize,
    frame=single,
    breaklines=true,
    showstringspaces=false,
    xleftmargin=0pt,
    showtabs=false,
    tabsize=2,
    belowcaptionskip=5pt,
    backgroundcolor=\color{gray!10}
}

\usepackage{tikz}
\usetikzlibrary{shapes.geometric, arrows}

\tikzstyle{startstop} = [rectangle, rounded corners, minimum width=3cm, minimum height=1cm,text centered, draw=black, fill=red!30]
\tikzstyle{process} = [rectangle, minimum width=3cm, minimum height=1cm, text centered, draw=black, fill=orange!30]
\tikzstyle{decision} = [diamond, minimum width=3cm, minimum height=1cm, text centered, draw=black, fill=green!30]
\tikzstyle{arrow} = [thick,->,>=stealth]

\usepackage{subcaption}

\newenvironment{answerbox}{
\begin{tcolorbox}[colback=blue!5!white,colframe=blue!5!white,arc=0mm,grow to left by=0mm,left=0mm,grow to right by=0mm,right=1.5mm,left=1.5mm,top=0mm,bottom=0mm]
}
{
\end{tcolorbox}
}

\AtBeginDocument{}

\begin{document}

\title[\methodName{}: Assembly-Level, Cross-Platform Testing of Quantum Computing Platforms]{\methodName{}: Assembly-Level, Cross-Platform Testing\\ of Quantum Computing Platforms}

\author{Matteo Paltenghi}
\orcid{0000-0003-2266-453X}
\affiliation{
  \institution{University of Stuttgart}
  \city{Stuttgart}
  \country{Germany}
}
\email{mattepalte@live.it}

\author{Michael Pradel}
\orcid{0000-0003-1623-498X}
\affiliation{
  \institution{CISPA Helmholtz Center for Information Security}
  \city{Stuttgart}
  \country{Germany}
}
\email{michael@binaervarianz.de}

\begin{abstract}
Quantum computing platforms are susceptible to quantum-specific bugs, such as incorrectly ordering qubits or incorrectly implementing quantum abstractions.
These bugs are difficult to detect and require specialized expertise. The field faces challenges due to a fragmented landscape of platforms and rapid development cycles that often prioritize new features over thorough testing, severely hindering the reliability of quantum software.
To address these challenges, we present \methodName{}, a novel cross-platform testing approach for quantum computing platforms.
The key technical contribution is our novel \coreComponentName{}, which generates equivalent quantum programs by iteratively (I)mporting them into platform-specific representations, (T)ransforming the program via optimizations and gate conversions, and (E)xporting the program again.
To transfer programs across platforms and test cross-platform consistency, \methodName{} leverages QASM, an assembly-level representation supported by most platforms.
The approach uses a crash oracle to detect failures during cross-platform transformations and an equivalence oracle to validate the semantic consistency of the generated assembly programs, which are expected to be equivalent by construction.
We evaluate \methodName{} on widely-used quantum computing platforms, including Qiskit, PennyLane, Pytket, BQSKit, and Cirq, revealing \TotalAllBugs{} bugs, \TotalConfirmedOrFixedBugs{} of which are already confirmed or fixed.
Our results also demonstrate that \methodName{} complements existing quantum fuzzers (covering tens of thousands of otherwise uncovered lines), is efficient (with \AvgGeneratorTimeOverall{} seconds per generated program), and that the \coreComponentName{} is crucial for its effectiveness.

\end{abstract}



\maketitle

\section{Introduction}

Quantum computing promises to solve computationally challenging problems beyond the reach of traditional computing.
An essential prerequisite to realize this potential is having reliable \emph{quantum computing platforms}, i.e., software systems that let developers define, manipulate, and execute quantum programs.
Growing interest from industry and research highlights the importance of these platforms for advancing quantum computing.
However, ensuring their reliability is challenging due to the complexity of quantum mechanics and the quantum software stack.
These challenges lead to quantum-specific bugs that require specialized expertise to detect and fix~\citep{paltenghiBugsQuantumComputing2022}.
Platform bugs can cause incorrect program behavior, impeding the development of quantum applications.
The intricate nature of these bugs can undermine trust in quantum computing platforms, motivating dedicated testing techniques to ensure reliability.

The quantum software landscape is characterized by a diverse ecosystem of platforms, including Qiskit~\cite{qiskit2024}, Cirq~\cite{QuantumlibCirq2021}, PennyLane~\cite{bergholmPennyLaneAutomaticDifferentiation2020}, Pytket~\cite{sivarajahKetRetargetableCompiler2020}, and BQSKit~\cite{younisBerkeleyQuantumSynthesis2021}, each providing a unique framework for quantum programming.
To ensure interoperability in this diverse ecosystem, many quantum computing platforms share a common representation for quantum programs: Quantum Assembly Language (QASM)~\cite{crossOpenQuantumAssembly2017}, which serves as a standard assembly language for representing quantum circuits.
Most quantum computing platforms support importing and exporting circuits in QASM format, making it a key enabler of cross-platform compatibility.

While the fragmentation of the quantum software ecosystem drives innovation, it introduces challenges in ensuring interoperability and causes gaps in current testing practices.
Existing approaches for testing quantum computing platforms~\cite{paltenghiMorphQMetamorphicTesting2023, paltenghiAnalyzingQuantumPrograms2024} are tailored to specific platforms, lacking the ability to validate the consistency of quantum circuit execution across different platforms.
This cross-platform consistency matters when developers exchange programs across specialized quantum computing platforms that target different architectures.
Therefore, a need for comprehensive cross-platform testing is emerging to ensure reliable interoperability across this diverse landscape.

\begin{figure}[t]
    \centering
    \begin{subfigure}{0.48\textwidth}
        \centering
        \begin{lstlisting}[style=coolpython]
from qiskit import QuantumCircuit, QuantumRegister
qr = QuantumRegister(4)
qc = QuantumCircuit(qr)
qc.mcx([qr[0], qr[1], qr[2]], qr[3])
        \end{lstlisting}
        \caption{Qiskit circuit with a three-control MCX.}
        \label{lst:bug_example_mcx_gate_py}
    \end{subfigure}
    \hfill
    \begin{subfigure}{0.45\textwidth}
        \centering
        \begin{lstlisting}[style=coolqasm]
OPENQASM 2.0;
include "qelib1.inc";
qreg q2[4];
c4x q2[0],q2[1],q2[2],q2[3];
        \end{lstlisting}
        \caption{Incorrect QASM from Pytket (\code{c4x} instead of \code{c3x}).}
        \label{lst:bug_example_mcx_gate_qasm}
    \end{subfigure}
    \hspace{2mm}
    \caption{Pytket generates an incorrect \code{c4x} gate for a three-control MCX operation (Bug~\IdBugIncorrectMcxGateGenerationCxInsteadOfCx{}).}
    \label{fig:bug_example_mcx_gate}
\end{figure}

To illustrate how bugs in a platform's assembly code generation and import/export functions can introduce subtle compatibility issues, consider Figure~\ref{fig:bug_example_mcx_gate}.
When converting and exporting a quantum program with a three-control multiple-controlled NOT operation (MCX), as shown in Listing~\ref{lst:bug_example_mcx_gate_py}, Pytket generates incorrect QASM assembly code.
The resulting QASM uses a \code{c4x} gate (for four controls) instead of the required \code{c3x} gate (for three controls), as shown in Listing~\ref{lst:bug_example_mcx_gate_qasm}.
This bug in Pytket's gate definition causes crashes when other platforms import the generated assembly code.
Unfortunately, the issue easily goes undetected when testing Pytket in isolation, since it can read back its own incorrect code.

The above bug illustrates a more general challenge:
While interoperability across quantum computing platforms is essential, there currently is no systematic approach to test this interoperability layer.
On the positive side, having multiple platforms that share a common assembly representation also offers opportunities:
Checking for cross-platform consistency can serve as a test oracle, which can help uncover subtle bugs in the assembly-level import/export functionality of quantum computing platforms.
Moreover, a cross-platform testing approach can test multiple platforms simultaneously, increasing the breadth of testing way beyond the current single-platform testing approaches~\cite{paltenghiMorphQMetamorphicTesting2023, paltenghiAnalyzingQuantumPrograms2024}.

This work introduces \methodName{}, a novel approach for cross-platform testing of quantum computing platforms.
\methodName{} harnesses the commonalities in quantum platforms through the use of a common assembly representation and employs an equivalence oracle to identify discrepancies in compiled circuit outputs.
\methodName{} starts by generating programs in the common assembly language.
For each generated program, \methodName{} then produces an equivalence class of semantically equivalent programs by iteratively importing, transforming, and exporting a program across different platforms under test.
This process, called the \emph{\coreComponentName{}}, is repeated multiple times to create a pool of semantically equivalent programs that can be used for cross-platform testing.
To amplify the number of equivalent programs, \methodName{} employs two transformation strategies: (a) applying platform-specific optimization passes and (b) converting the program gate set to a different but equivalent gate set.
Finally, \methodName{} uses two oracles to detect bugs: a \emph{crash oracle} that checks for crashes during all components of \methodName{} and an \emph{equivalence oracle} that compares the programs in the same equivalence class for semantic equivalence.
The approach is designed to test several critical components of quantum computing platforms, including the import and export of assembly programs, the internal representation of circuits, and the application of circuit transformations.
In contrast, testing the execution of quantum programs on quantum hardware is out of scope for \methodName{}.

Our evaluation shows that \methodName{} is highly effective at finding bugs in quantum computing platforms. Across five platforms, \methodName{} discovered \TotalAllBugs{} bugs, with \TotalConfirmedOrFixedBugs{} already confirmed or fixed by developers. These bugs span a range of issues, from incorrect gate implementations to compiler optimizations that fail to preserve program semantics. Moreover, by focusing on the interoperability layer, \methodName{} reveals unique issues that are typically out of scope for existing quantum fuzzers focused on a single platform, demonstrating strong complementarity with current testing approaches.

Our work advances the state-of-the-art in testing quantum computing platforms.
While some prior work, e.g., QDiff~\cite{wangQDiffDifferentialTesting2021}, MorphQ~\cite{paltenghiMorphQMetamorphicTesting2023}, Fuzz4All~\cite{xiaFuzz4AllUniversalFuzzing2024}, and QuteFuzz~\cite{iwumbweQuteFuzzFuzzingQuantum}, focuses on testing a single platform across different configurations and optimization levels, we introduce a more general approach that tests interoperability across multiple quantum computing platforms by leveraging the common assembly-level representation.
Other existing work, e.g., FuzzQ~\cite{klimis2025shaking} and work by Blackwell et al.~\cite{blackwellFuzzingBasedDifferentialTesting2024}, also perform cross-platform differential testing, but compare simulator behavior instead of testing cross-platform interoperability.
A key difference to all the above approaches is that \methodName{} is the only one that uses cross-platform interoperability as a means to generate new quantum programs.

The contributions of this work include:
\begin{itemize}
\item We introduce \methodName{}, a novel approach for cross-platform testing of quantum computing platforms.
\item We present the novel ITE process, which iteratively imports, transforms, and exports programs across different platforms to create a pool of semantically equivalent programs.
\item We conduct an extensive evaluation of \methodName{} on five widely-used quantum computing platforms, demonstrating its effectiveness in detecting \TotalAllBugs{} bugs, with (so far) \TotalConfirmedOrFixedBugs{} confirmed and/or fixed by developers, and showing its complementarity to existing quantum fuzzers.
\item We provide code and data (Section~\ref{sec:data_availability}) to reproduce our results, enhancing transparency and inspiring future research.
\end{itemize}

\section{Background}
\subsection{Quantum Software and QASM}
Quantum computing platforms enable developers to define, manipulate, and execute quantum programs. These programs take the form of quantum circuits, where register definitions first specify the available qubits, followed by a sequence of quantum operations (gates) applied to specific qubits or groups of qubits. Through these structured programs, developers implement quantum algorithms, with the platform managing the complex task of preparing the programs for execution on quantum hardware.

\begin{figure}[t]
    \centering
    \begin{subfigure}{0.45\textwidth}
        \centering
        \begin{lstlisting}[style=qasm, basicstyle=\ttfamily\footnotesize, escapeinside={(*}{*)}]
OPENQASM 2.0;
include "qelib1.inc";
qreg q[2];
(*\color{blue}{h} q[0]*);
(*\color{darkgreen}{swap} q[0],q[1]*);

(*~*)
        \end{lstlisting}
        \caption{Original QASM.}
        \label{fig:qasm_example}
    \end{subfigure}
    \hfill
    \begin{subfigure}{0.45\textwidth}
        \centering
        \begin{lstlisting}[style=qasm, basicstyle=\ttfamily\footnotesize, escapeinside={(*}{*)}]
OPENQASM 2.0;
include "qelib1.inc";
qreg q[2];
(*\color{blue}{u3}(pi/2,0,pi) q[0]*);
(*\color{darkgreen}{cx} q[0],q[1]*);
(*\color{darkgreen}{cx} q[1],q[0]*);
(*\color{darkgreen}{cx} q[0],q[1]*);
        \end{lstlisting}
        \caption{Transformed QASM.}
        \label{fig:qasm_example_transformation}
    \end{subfigure}
    \hspace{2mm}

    \vspace{0.2cm}

    \begin{subfigure}{0.45\textwidth}
        \centering
        \scalebox{0.7}{
        \begin{quantikz}
            \lstick{$q_0$} & \gate{H} & \swap{1} & \qw \\
            \lstick{$q_1$} & \qw & \swap{-1} & \qw
        \end{quantikz}
        }
        \caption{Original circuit.}
        \label{fig:qasm_example_circuit}
    \end{subfigure}
    \hfill
    \begin{subfigure}{0.45\textwidth}
        \centering
        \scalebox{0.7}{
        \begin{quantikz}
            \lstick{$q_0$} & \gate{U_3(\pi/2,0,\pi)} & \ctrl{1} & \targ{} & \ctrl{1} & \qw \\
            \lstick{$q_1$} & \qw & \targ{} & \ctrl{-1} & \targ{} & \qw
        \end{quantikz}
        }
        \caption{Transformed circuit.}
        \label{fig:qasm_example_transformed_circuit}
    \end{subfigure}

    \caption{Original quantum circuit and its transformed version using the U3 and CX gate set.}
\end{figure}

QASM~\cite{crossOpenQuantumAssembly2017} serves as a standard assembly-level representation to specify quantum programs.
Like traditional assembly languages, QASM follows a simple structure.
Each program starts with a prologue that declares the version and includes standard libraries.
The program body consists of register definitions, analogous to variable declarations in classical programs, followed by quantum operations.
Each operation specifies a gate name, optional parameters, and the qubits it acts on.
Figure~\ref{fig:qasm_example} illustrates this structure with a simple quantum program.
The program applies a Hadamard gate (\code{h}) to qubit 0 -- a quantum operation that puts the qubit in an equal superposition of the states 0 and 1 -- followed by a swap operation (\code{swap}) between qubits 0 and 1 -- a quantum operation that swaps the states of the two qubits.
Figure~\ref{fig:qasm_example_circuit} shows the corresponding quantum circuit diagram.

\subsection{Program Transformations}
Program transformations are essential techniques in quantum computing for generating semantically equivalent quantum programs. Two fundamental types of transformations are optimization passes and gate set conversions.
Optimization passes reduce circuit complexity by minimizing the number of quantum gates while preserving behavior.
Gate set conversions transform circuits to use different sets of quantum gates, which is particularly useful when targeting specific hardware platforms.
As illustrated in Figures~\ref{fig:qasm_example_transformation} and~\ref{fig:qasm_example_transformed_circuit}, a quantum circuit can be transformed into an equivalent version using only \code{u3} and \code{cx} gates, a universal gate set commonly used in quantum computing platforms like Qiskit~\cite{qiskit2024}.

\section{The \methodName{} Approach}
\label{sec:approach}

We start by defining the problem this paper addresses, and then present the components of \methodName{}.

\subsection{Problem Statement}
\label{sec:problem_statement}

We address the challenge of cross-platform testing for quantum computing platforms.
Given a set of $n$ quantum computing platforms $P = \{p_1, \ldots, p_n\}$, we assume that each platform $p$ provides an import function $i$ and an export function $e$ that support a common assembly language shared by all platforms, e.g., QASM~\citep{crossOpenQuantumAssembly2017}.
Let $A$ be the set of assembly programs and $R$ be the set of platform-specific representations.
The importer function $i_p: A \rightarrow R$ converts an assembly program into the platform's representation.
The exporter function $e_p: R \rightarrow A$ converts the platform's representation back into assembly.
Additionally, each platform can offer transformations $t \in T_p$ that produce semantically equivalent programs, i.e., $t: R \rightarrow R$ and for any $r, r' \in R$, if $r' = t(r)$ then $r \sim r'$, where $\sim$ denotes semantic equivalence.
Note that \methodName{} also supports platforms that do not offer any transformations, i.e., $T_p = \emptyset$.

Our testing goal is to automatically detect two types of bugs:
(a) platform crashes that occur while a platform manipulates programs, e.g., while running $i_p$, $e_p$, or any $t \in T_p$, and
(b) semantic inequivalences $a \nsim a'$ between assembly programs that, by construction, should be equivalent.

\subsection{Overview}
\label{sec:overview}

\begin{figure*}[t]
    \centering
    \includegraphics[width=.9\textwidth]{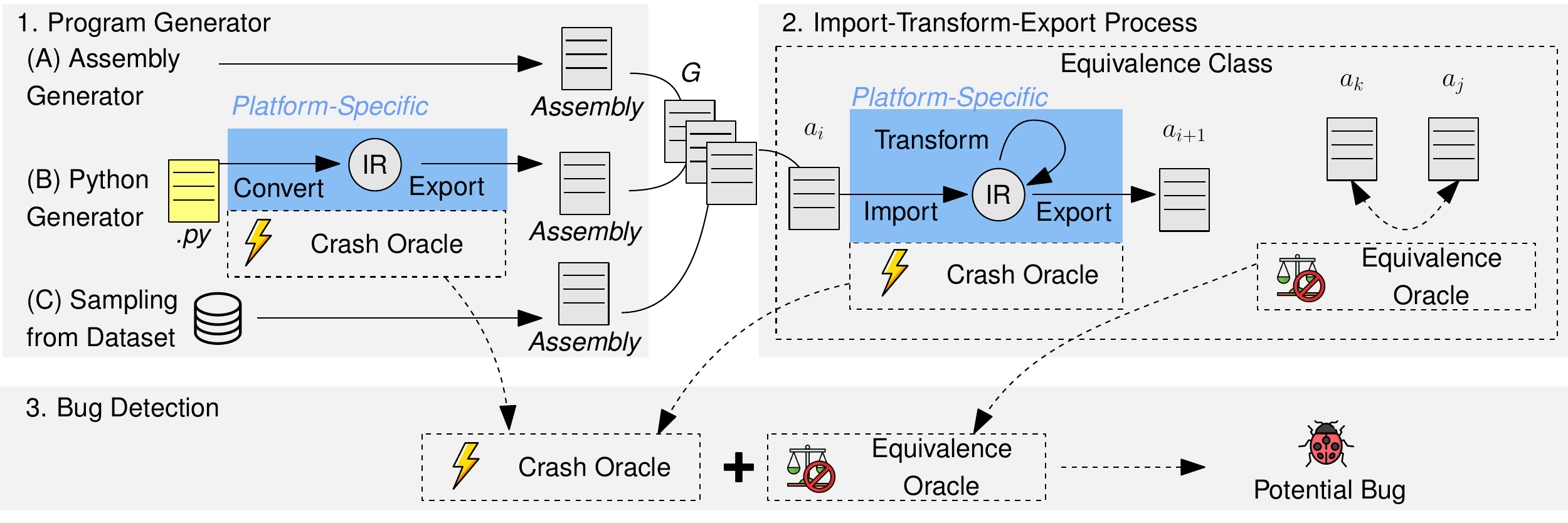}
    \caption{Overview of \methodName{}. The platform-specific components (blue boxes) are implemented by randomly sampling from the platforms under test.}
    \label{fig:overview_cross_platform_testing}
  \end{figure*}

To address this goal, \methodName{} uses a three-component approach: program generation, our novel \coreComponentName{}, and bug detection (Figure~\ref{fig:overview_cross_platform_testing}).
First, the program generation component (Sec.~\ref{sec:program_generation}) produces quantum assembly programs either by directly generating assembly code, by transforming platform-specific programs via platform converters, or by sampling from a pre-existing corpus of programs.
Second, the \coreComponentName{} (Sec.~\ref{sec:core_component}) iteratively imports, transforms, and exports programs across platforms.
This process creates new programs that form an equivalence class, because importing, transforming, and exporting are supposed to preserve semantic equivalence.
An equivalence class $E$ contains programs such that $a_i \sim a_j$ for all $a_i, a_j \in E$.
Each conversion and transformation is platform-specific, relying on the internal code of the platform under test.
Finally, the bug detection component (Sec.~\ref{sec:bug_detection}) uses two oracles: a crash oracle that monitors program generation and the \coreComponentName{}, and an equivalence oracle that checks semantic equivalence within each equivalence class.
If either oracle detects a crash or an inequivalence, \methodName{} reports a bug.

\subsection{Program Generation}
\label{sec:program_generation}

To create quantum programs, \methodName{} supports three generators that produce assembly code in QASM.
The first two generators build upon a template-based approach, which has proven effective in uncovering bugs when testing a single quantum computing platform~\cite{paltenghiMorphQMetamorphicTesting2023}, whereas the third generator samples program from an existing corpus of quantum programs.

The first generator, referred to as the \emph{assembly-based generator}, creates QASM code directly using a grammar-based approach, as shown in Figure~\ref{fig:quantum_grammar}. The grammar defines the basic structure of quantum programs, including register declarations and quantum operations, ensuring that the generated programs are syntactically valid.
One example of possible QASM code generated by this generator is shown in Figure~\ref{fig:qasm_example}.

\begin{figure}[t]
    \centering
    \begin{subfigure}{.97\textwidth}
        \centering
        \begin{lstlisting}[style=coolpython]
from pytket.extensions.qiskit import qiskit_to_tk
pytket_circ = qiskit_to_tk(qc)
        \end{lstlisting}
        \caption{Conversion from Qiskit to Pytket representation.}
        \label{fig:pytket_conversion}
    \end{subfigure}
    \hspace{2mm}

    \vspace{.3cm}

    \begin{subfigure}{.97\textwidth}
        \centering
        \begin{lstlisting}[style=coolpython]
from pytket.qasm import circuit_to_qasm_str
qasm_str = circuit_to_qasm_str(pytket_circ)
        \end{lstlisting}
        \caption{Exporter to QASM in Pytket.}
        \label{fig:pytket_exporter}
    \end{subfigure}
    \hspace{2mm}

    \caption{Pytket's platform-specific converter and exporter APIs.}
    \label{fig:pytket_apis}
\end{figure}

The second generator, referred to as the \emph{Python-based generator}, produces high-level quantum programs using Qiskit's Python API~\cite{qiskit2024}. The grammar closely resembles the QASM grammar since the Qiskit API mirrors QASM syntax. An example of a Qiskit program generated by the Python-based generator is shown in Listing~\ref{lst:bug_example_mcx_gate_py}.
Many quantum platforms provide converters from Qiskit to their internal representations. This generator leverages these platform-specific converters to transform the generated Qiskit code into other platforms' representations. Figure~\ref{fig:pytket_conversion} shows an example of conversion from a Qiskit circuit to Pytket's internal representation using platform-specific APIs. The internal representation is then exported to QASM using another platform-specific API, e.g., as shown in Figure~\ref{fig:pytket_exporter}.
\methodName{} supports multiple platforms, each with a different converter and specific process to convert the Qiskit representation to the platform's internal representation and back to QASM. Both the converter code and the subsequent export to QASM are implemented by the platform maintainers, ensuring that we test only the platform's code during this translation process.
We select Qiskit to generate the initial Python program because it was developed by the same company that proposed the QASM specification, IBM. This alignment ensures that Qiskit offers the most robust support for QASM features during translation. Our evaluation confirms this, as Qiskit's implementation of the QASM specification stands out as being particularly robust.

The third generator, referred to as \emph{sampling from dataset}, retrieves programs from a pre-existing corpus of diverse quantum programs~\cite{quetschlichMQTBenchBenchmarking2023}.
This corpus contains implementations of a variety of quantum algorithms, and hence, complements the other two generators by offering realistic programs.

\methodName{} can be configured to use any subset of these generators, allowing for flexibility in program generation.
By default, the approach uses all three generators equally, alternating between them to produce batches of programs.
For the third generator, \methodName{} samples each program at most once to maximize diversity, falling back to the other generators once the pre-existing corpus is exhausted.
Our evaluation studies the impact of the generators in detail (Section~\ref{sec:ablation study}).

\begin{figure}[t]
\centering
\small
\begin{tabular}{|lcl|}
\hline
\textcolor{blue}{Program} & ::= & \textcolor{blue}{Header} \textcolor{blue}{RegisterDecl} \textcolor{blue}{StatementList} \\
\textcolor{blue}{Header} & ::= & \code{"OPENQASM 2.0;"} \\
\textcolor{blue}{RegisterDecl} & ::= & \code{"qreg q["} \textcolor{blue}{QubitCount} \code{"];" } \\
            &     & \code{"creg c["} \textcolor{blue}{QubitCount} \code{"];" } \\
\textcolor{blue}{StatementList} & ::= & \textcolor{blue}{Statement}* \\
\textcolor{blue}{Statement} & ::= & \textcolor{blue}{GateStatement} \code{";"} \\
\textcolor{blue}{GateStatement} & ::= & \textcolor{blue}{SingleQubitGate} $|$ \textcolor{blue}{TwoQubitGate} $|$ ...\\
\textcolor{blue}{SingleQubitGate} & ::= & \code{"h q["} \textcolor{blue}{QIx} \code{"]"} $|$ \code{"rx("} \textcolor{blue}{Param} \code{") q["} \textcolor{blue}{QIx} \code{"]"} $|$ ...\\
\textcolor{blue}{TwoQubitGate} & ::= & \code{"cx q["} \textcolor{blue}{QIx} \code{"],q["} \textcolor{blue}{QIx} \code{"]"} $|$ \code{"crz("} \textcolor{blue}{Param} \code{") q["} \textcolor{blue}{QIx} \code{"],q["} \textcolor{blue}{QIx} \code{"]"} $|$ ... \\[1ex]
\multicolumn{3}{|l|}{Constraints:} \\[1ex]
\multicolumn{3}{|l|}{$\bullet$ \textcolor{blue}{QubitCount}, \textcolor{blue}{QIx} $\in \mathbb{N}$ $\wedge$ \textcolor{blue}{Param} $\in \mathbb{R}$} \\[1ex]
\multicolumn{3}{|l|}{$\bullet$ \textcolor{blue}{QIx} $<$ \textcolor{blue}{QubitCount} $\wedge$ \textcolor{blue}{QubitCount}, \textcolor{blue}{QIx} $\geq 0$} \\
\hline
\end{tabular}
\caption{Grammar of our quantum assembly generator, defining the structure of quantum programs with register declarations and quantum operations. The \texttt{h} and \texttt{rx} gates act on single qubits, while \texttt{cx} and \texttt{crz} gates act on pairs of qubits. More gates are supported but omitted for brevity.}
\label{fig:quantum_grammar}
\end{figure}

\subsection{Import--Transform--Export Loop}
\label{sec:core_component}

\begin{algorithm}[t]
\caption{Import-Transform-Export (ITE) process}
\label{alg:qite_process}
\begin{algorithmic}[1]
\Require
    \State $P$: Quantum platforms $\{p_1, \ldots, p_n\}$
    \State $G$: Set of generated programs
    \State $T_p$: Transformations for platform $p \in P$
    \State $m$: Number of ITE iterations
\Ensure
    \State $EQ$: Set of equivalence classes of assembly programs
    \State $EQ \gets \emptyset$ \label{line:init_eq} \Comment{Initialize the set of equivalence classes}
    \While {testing budget remaining}
        \State $a \gets \text{Sample}(G)$ \label{line:pick_initial_program} \Comment{Pick initial assembly program}
        \State $E \gets \{a\}$ \label{line:init_eq_class} \Comment{Start a new equivalence class}
        \For {$i \gets 1$ \textbf{to} $m$} \label{line:transformation_loop} \Comment{Iterate through ITE iterations}
            \State $p \gets \text{Sample}(P)$ \label{line:sample_platform} \Comment{Select a platform}
            \State $r \gets i_p(a)$ \label{line:import_qasm} \Comment{Import assembly into platform $p$}
            \State $t \gets \text{Sample}(T_p)$ \label{line:sample_transformation} \Comment{Select a transform for platform $p$}
            \State $r' \gets t(r)$ \label{line:apply_transformation} \Comment{Apply transformation $t$}
            \State $a' \gets e_p(r')$ \label{line:export_qasm} \Comment{Export to assembly}
            \State $E.\text{append}(a')$ \label{line:append_eq_class} \Comment{Add to equivalence class}
            \State $a \gets a'$ \label{line:update_current_program} \Comment{Update current program}
        \EndFor
        \State $EQ.\text{append}(E)$ \label{line:append_eq} \Comment{Add equivalence class to list}
    \EndWhile
    \State \Return $EQ$ \label{line:return_eq} \Comment{Return list of equivalence classes}
\end{algorithmic}
\end{algorithm}

The \coreComponentName{} is the central component of \methodName{} and our key technical contribution.
Algorithm~\ref{alg:qite_process} shows the pseudocode for this iterative process. In each iteration, the algorithm imports the generated assembly code into a specific platform, transforms it, and then exports it back to assembly.  The algorithm picks an initial assembly program $a$ from the set of generated programs $G$ (line~\ref{line:pick_initial_program}) and initializes an equivalence class $E$ with $a$ (line~\ref{line:init_eq_class}).
An example of an input assembly program is shown in Figure~\ref{fig:qasm_example}.
Then, the algorithm iterates $m$ times (line~\ref{line:transformation_loop}). Within this loop, it samples a platform $p$ from the set of platforms $P$ (line~\ref{line:sample_platform}). The algorithm imports the current assembly program $a$ into the platform $p$ (line~\ref{line:import_qasm}) converting it to the platform-specific representation $r$. The \coreComponentName{} then samples a transformation $t$ (line~\ref{line:sample_transformation}) and applies it to $r$ to obtain $r'$ (line~\ref{line:apply_transformation}), which is a semantically equivalent representation of the program $a$.
Next, the algorithm exports $r'$ back to assembly, yielding the transformed assembly program $a'$ (line~\ref{line:export_qasm}).
An example of an output assembly program, after applying a transformation that changes its gate set, is shown in Figure~\ref{fig:qasm_example_transformation}.
The transformed assembly program $a'$ is added to the equivalence class $E$ (line~\ref{line:append_eq_class}), and the current program $a$ is updated to $a'$ (line~\ref{line:update_current_program}) for the next iteration.
After completing all $m$ ITE iterations, the equivalence class $E$ is added to the set of equivalence classes $EQ$ (line~\ref{line:append_eq}).
Finally, the algorithm returns the set of equivalence classes $EQ$ (line~\ref{line:return_eq}), where each equivalence class contains a set of semantically equivalent assembly programs.
The default value for $m$ is set to 5, but our evaluation (Section~\ref{sec:ablation study}) explores the impact of varying this parameter.

\begin{table}[t]
    \centering
    \caption{Optimization and gate set change APIs for different platforms.}
    \renewcommand{\arraystretch}{0.8}
    \begin{tabular}{p{1.5cm} p{5cm} p{5cm}}
        \toprule
        \textbf{Platform} & \textbf{Optimizations} & \textbf{Change Gate Set} \\
        \midrule
        \multirow{5}{*}{\rotatebox{90}{Qiskit}} &
        \scode{transpile(opt\_level=0)} & \multirow{1}{*}{\scode{transpile(basis\_gates=..)}} \\
        & \scode{transpile(opt\_level=1)} & \multirow{1}{*}{\scode{['u1', 'u2', 'u3', 'cx']}} \\
        & \scode{transpile(opt\_level=2)} & \multirow{1}{*}{\scode{['u3', 'cx']}} \\
        & \scode{transpile(opt\_level=3)} & \multirow{1}{*}{\scode{['u3', 'cx']}} \\
        &  & \multirow{1}{*}{\scode{['rz', 'sx', 'x', 'cx']}} \\
        &  & \multirow{1}{*}{\scode{['rx', 'ry', 'rz', 'cz']}} \\   \midrule   \multirow{4}{*}{\rotatebox{90}{Pytket}} &
        \scode{FullPeepholeOptimise} & \multirow{1}{*}{\scode{AutoRebase(gateset=..)}} \\
        & \scode{PeepholeOptimise2Q} & \scode{[OpType.U1, U2, U3, CX]}\\
        & \scode{RemoveRedundancies} & \scode{[OpType.U3, CX]}\\
        & \scode{OptimisePhaseGadgets} & \scode{[OpType.RZ, SX, X, CX]}\\
        &  & \scode{[OpType.RX, RY, RZ, CZ]}\\
        \midrule
        \multirow{11}{*}{\rotatebox{90}{PennyLane}} &
        \scode{cancel\_inverses} & \multirow{1}{*}{\scode{decompose}} \\
        & \scode{commute\_controlled} & \\
        & \scode{merge\_rotations} & \\
        & \scode{single\_qubit\_fusion} & \\
        & \scode{unitary\_to\_rot} & \\
        & \scode{remove\_barrier} & \\
        & \scode{undo\_swaps} & \\
        & \scode{combine\_global\_phases} & \\
        & \scode{clifford\_t\_decomposition} & \\
        & \scode{merge\_amplitude\_embedding} & \\
        & \scode{defer\_measurements} & \\
        \midrule
        \multirow{4}{*}{\rotatebox{90}{BQSKit}} &
        \scode{compile( opt\_level=0 )} & \multirow{4}{*}{N/A} \\
        & \scode{compile( opt\_level=1 )} & \\
        & \scode{compile( opt\_level=2 )} & \\
        & \scode{compile( opt\_level=3 )} & \\
        \midrule
        \multirow{6}{*}{\rotatebox{90}{Cirq}} &
        \scode{drop\_empty\_moments} & \multirow{6}{*}{N/A} \\
        & \scode{eject\_z} & \\
        & \scode{eject\_phased\_paulis} & \\
        & \scode{align\_left} & \\
        & \scode{align\_right} & \\
        & \scode{drop\_negligible\_operations} & \\
        \bottomrule
    \end{tabular}
    \label{tab:optimization_gateset}
\end{table}

Table~\ref{tab:optimization_gateset} summarizes the transformations $T_p$ available for the five platforms supported by \methodName{}: Qiskit, Pytket, PennyLane, BQSKit, and Cirq. To ensure semantic equivalence, the \coreComponentName{} employs two main types of transformations: optimization passes and gate set conversions.
Each transformation takes a circuit in the platform's internal representation and returns a semantically equivalent circuit in the same representation.
Optimization passes aim to reduce the number of gates in the circuit, while gate set conversions adapt the circuit to a different gate set.
As there are (theoretically) infinitely many universal gate sets, we select a few commonly used ones for gate set conversions, such as \code{['u3', 'cx']} and \code{['rz', 'sx', 'x', 'cx']}.
Using commonly used gate sets means that \methodName{} tests widely used transformations and ensures that a program transformed by one platform can be imported and transformed by other platforms.
Importantly, all these transformations are supposed to be semantics-preserving, i.e., they do not change the behavior of the program.

For Qiskit \citep{qiskit2024}, \methodName{} uses the \code{transpile} API to apply transformations such as optimizing the circuit at different levels (e.g., \code{opt\_level=1}) or changing the gate set to a universal gate set composed of \code{['u3', 'cx']} gates, an example of the effect of which is shown in Figure~\ref{fig:qasm_example_transformation}.
For PennyLane \citep{bergholmPennyLaneAutomaticDifferentiation2020}, \methodName{} uses circuit transformations such as \code{cancel\_inverses} to simplify the circuit by canceling inverse gates and \code{merge\_rotations} to combine consecutive rotation gates.
For Pytket \citep{sivarajahKetRetargetableCompiler2020}, \methodName{} uses optimization techniques such as \code{FullPeepholeOptimise} to perform peephole optimization and \code{AutoRebase} to perform gate set transformations.
For BQSKit, \methodName{} leverages the predefined optimization levels offered by the \code{compile} function, e.g., \code{compile(opt\_level=2)}.
For Cirq~\cite{QuantumlibCirq2021}, \methodName{} wraps built-in circuit transformers such as \code{drop\_empty\_moments}, \code{eject\_z}, \code{eject\_phased\_paulis}, \code{align\_left}, \code{align\_right}, and \code{drop\_negligible\_operations}. Since Cirq's QASM importer has stricter dialect assumptions than some other platforms, the Cirq processor also applies lightweight QASM normalization before import, including rewriting Qiskit-style \code{u(...)} operations to \code{u3(...)}.

\methodName{}'s design allows for easy extensibility to new quantum computing platforms.
To apply the approach to a new platform, the platform must provide three key components: an importer, an exporter, and optionally, one or more transformations.

\subsection{Bug Detection}
\label{sec:bug_detection}

The bug detection component is active during all the previous stages of \methodName{}.
It employs two complementary oracles to detect potential bugs.
The \emph{crash oracle} is always active, monitoring the platform code for crashes.
Specifically, crashes may occur when the generator uses a platform's converter and when the \coreComponentName{} invokes a platform's importer, transformations, and exporter.
The \emph{equivalence oracle} is periodically triggered when the \coreComponentName{} completes a given number of ITE iterations on a given set of initial assembly programs.
The oracle is based on QCEC~\citep{pehamEquivalenceCheckingQuantum2022}, an existing equivalence checker for quantum circuits.
The checker gets the set of equivalence classes $EQ$ generated by the \coreComponentName{} and compares pairs of programs within each equivalence class $E$ to ensure semantic equivalence.
Since all programs in $E$ are constructed to be semantically equivalent, any discrepancies detected by the equivalence checker signal a bug.
We call each pair of programs flagged as not semantically equivalent by the equivalence oracle an \emph{inequivalence}.
To address the scalability issue of comparing all pairs of programs within an equivalence class, \methodName{} randomly samples a small subset of program pairs for comparison (by default, five pairs per equivalence class).

Our equivalence oracle differs from differential testing oracles that compare program behavior across different optimization levels or simulator backends~\cite{wangQDiffDifferentialTesting2021,iwumbweQuteFuzzFuzzingQuantum,klimis2025shaking} because it does not execute the programs.
Instead, it uses the symbolic equivalence-checking method of QCEC, which translates the two quantum circuits into ZX-diagrams and symbolically checks whether the composition of one circuit's inverse with the other simplifies to the identity~\citep{pehamEquivalenceCheckingQuantum2022}.
The advantage of this approach is that it can detect semantic discrepancies without relying on specific input states or measurement outcomes, which are inherently probabilistic in quantum computing and can lead to false positives or negatives in differential testing.

To streamline the inspection of detected bugs, the approach heuristically reduces bug-triggering programs.
When \methodName{} detects a potential issue, it records the exception or the specific discrepancy reported by the equivalence checker. Then, \methodName{} uses delta debugging~\cite{zellerIsolatingCauseeffectChains2002} to reduce the program to the smallest test case that still triggers the issue.
Specifically, \methodName{} uses the exception message as the signal for crashes and the returned string \code{not equivalent} reported by the equivalence checker as the signal for inequivalences.
To ensure reproducibility and allow for automated delta debugging, \methodName{}'s metadata tracks the provenance of each generated program, including the sequence of transformations applied during the \coreComponentName{}.
This allows us to replay the exact steps leading to a bug.
While the heuristic reduction cannot guarantee that the final reduced program (or pair of reduced programs) preserves the original root cause, a minimal program (or pair of programs) that triggers a crash or inequivalence remains valuable to developers.

To prioritize bug inspection and reduce redundancy, \methodName{} groups warnings based on their error messages, assuming that identical messages indicate the same underlying bug. Then, we sample and inspect one warning per cluster, representing a unique error message. In the case of equivalence discrepancies, where no explicit error message is available, we sample and inspect a subset of the generated programs.

\section{Evaluation}
\label{sec:evaluation}

We investigate the following research questions:
\begin{itemize}
  \item \textbf{RQ1: Bug detection}. How effective is \methodName{} at finding bugs in quantum computing platforms?
  \item \textbf{RQ2: Impact of ITE iterations}. What is the impact of the number $m$ of ITE iterations on the generated programs and the number of detected crashes and inconsistencies?
  \item \textbf{RQ3: Efficiency}. How efficient are the different components of \methodName{}?
  \item \textbf{RQ4: Ablation study}. How does the choice of program generators and transformations affect the effectiveness of \methodName{}?
  \item \textbf{RQ5: Comparison with prior work}. How does \methodName{} compare with prior tools for testing quantum computing platforms?
\end{itemize}

\subsection{Experimental Setup}

We evaluate \methodName{} on five widely-used quantum computing platforms: Qiskit~\cite{qiskit2024} (Qiskit version 1.2.4, Qiskit Aer version 0.15.1, and Qiskit IBM Runtime version 0.29.0), PennyLane~\cite{bergholmPennyLaneAutomaticDifferentiation2020} (PennyLane version 0.40.0 and PennyLane-qiskit version 0.40.0), Pytket~\cite{sivarajahKetRetargetableCompiler2020} (Pytket version 1.33.1 and Pytket-qiskit version 0.56.0), BQSKit~\cite{younisBerkeleyQuantumSynthesis2021} (version 1.2.0), and Cirq~\cite{QuantumlibCirq2021} (version 1.5.0).
While we consider the BQSKit platform for RQ1 to assess the bug-finding effectiveness of \methodName{}, we exclude it from subsequent research questions because its exporter produces many invalid assembly programs due to a single bug, which disrupts the continuous application of ITE iterations.
Moreover, its execution often reaches timeout thresholds, likely due to its pure Python implementation without performance optimizations in systems languages, unlike Qiskit's Rust or Pytket's C++ components.
For the equivalence oracle, we use MQT QCEC~\citep{pehamEquivalenceCheckingQuantum2022} version 2.7.1.
We conduct all experiments on a dedicated machine equipped with 48 CPU cores (Intel Xeon Silver, 2.20GHz), two NVIDIA Tesla T4 GPUs with 16GB of memory each, and 252GB of RAM, running Ubuntu 22.04.5 LTS.

\subsection{RQ1: Bug Detection}

We evaluate the effectiveness of \methodName{} in detecting crashes and inconsistencies across five quantum computing platforms: Qiskit, PennyLane, Pytket, BQSKit, and Cirq.
We measure \methodName{}'s effectiveness by the number of unique bugs found, the number confirmed and fixed by developers, and developer feedback on GitHub issues we create.
The results for this RQ are from several runs of \methodName{} performed over a period of three months.

Table~\ref{tab:bugs_summary} summarizes the bugs found across the different platforms.
For each bug, we provide the platform and GitHub issue number, a brief description, its current status, and the oracle (crash or equivalence) that detected it.
The results show that \methodName{} effectively identifies bugs, uncovering \TotalAllBugs{} issues across the evaluated platforms, with \TotalConfirmedOrFixedBugs{} already confirmed or fixed, demonstrating its practical impact and the interest by platform maintainers to address these issues.

\begin{table*}[t]
    \centering
\caption{Summary of bugs found in different platforms.}
\label{tab:bugs_summary}
\setlength{\tabcolsep}{2pt}
\begin{tabular}{@{}llllll@{}}
\toprule
Id & Issue & Platform & Short description & Status & Oracle \\
\midrule
\#1 & \href{https://github.com/PennyLaneAI/pennylane/issues/6323}{\#6323} & PennyLane & Unexpected reallocation of gates to different qubits & Confirmed & Equiv. \\
\#2 & \href{https://github.com/Qiskit/qiskit/issues/12124}{\#12124} & Qiskit & Missing standard \code{rxx} gate definition & Confirmed & Crash \\
\#3 & \href{https://github.com/BQSKit/bqskit/issues/307}{\#307} & BQSKit & Unrecognized delay gate & Confirmed & Crash \\
\#4 & \href{https://github.com/CQCL/tket/issues/1629}{\#1629} & Pytket & Error converting decomposed circuit (\code{PhasedX} op) & Confirmed & Crash \\
\#5 & \href{https://github.com/BQSKit/bqskit/issues/306}{\#306} & BQSKit & Missing controlled phase \code{cs} gate definition & Confirmed & Crash \\
\#6 & \href{https://github.com/PennyLaneAI/pennylane/issues/6900}{\#6900} & PennyLane & Error exporting QASM file: \code{MidMeasureMP} op & Fixed & Crash \\
\#7 & \href{https://github.com/PennyLaneAI/pennylane/issues/6942}{\#6942} & PennyLane & Unitary matrix mismatch in \code{cu} gate implementation & Confirmed & Equiv. \\
\#8 & \href{https://github.com/CQCL/tket/issues/1771}{\#1771} & Pytket & Incompat. of \code{PhasedX} gate with \code{GreedyPauliSimp} opt & Confirmed & Equiv. \\
\#9 & \href{https://github.com/Quantinuum/tket/issues/1652}{\#1652} & Pytket & QASM register reordering causes semantic changes & Confirmed & Equiv. \\
\#10 & \href{https://github.com/CQCL/tket/issues/1605}{\#1605} & Pytket & Unsupported \code{ISwap} gate & Fixed & Crash \\
\#11 & \href{https://github.com/CQCL/tket/issues/1747}{\#1747} & Pytket & \code{ZXGraphlikeOpt} pass creates non-equivalent circuit & Fixed & Equiv. \\
\#12 & \href{https://github.com/CQCL/tket/issues/1750}{\#1750} & Pytket & \code{FullPeepholeOpt} pass fails with \code{mcrz} gate & Fixed & Crash \\
\#13 & \href{https://github.com/CQCL/tket/issues/1751}{\#1751} & Pytket & Incorrect \code{MCX} gate generation (\code{c4x} instead of \code{c3x}) & Fixed & Crash \\
\#14 & \href{https://github.com/CQCL/tket/issues/1768}{\#1768} & Pytket & Incorrect \code{u0} gate conversion to \code{u3} & Fixed & Crash \\
\#15 & \href{https://github.com/CQCL/tket/issues/1770}{\#1770} & Pytket & \code{FullPeepholeOpt} pass fails with \code{u0} and \code{c4x} gates & Fixed & Crash \\
\#16 & \href{https://github.com/BQSKit/bqskit/issues/302}{\#302} & BQSKit & Duplicate classical register definition & Reported & Crash \\
\#17 & \href{https://github.com/BQSKit/bqskit/issues/298}{\#298} & BQSKit & Crash with DaggerGate & Reported & Crash \\
\#18 & \href{https://github.com/BQSKit/bqskit/issues/299}{\#299} & BQSKit & Missing Ryy gate definition & Reported & Crash \\
\#19 & \href{https://github.com/quantumlib/Cirq/issues/8087}{\#8087} & Cirq & QASM importer rejects lowercase \code{u} gate & Fixed & Crash \\
\#20 & \href{https://github.com/PennyLaneAI/pennylane/issues/9499}{\#9499} & PennyLane & \code{merge\_amplitude\_embedding} inequivalence & Reported & Equiv. \\
\#21 & \href{https://github.com/Quantinuum/tket/issues/2167}{\#2167} & Pytket & \code{GreedyPauliSimp} changes circuit semantics & Confirmed & Equiv. \\
\#22 & \href{https://github.com/Quantinuum/tket/issues/2169}{\#2169} & Pytket & \code{PeepholeOptimise2Q} changes circuit semantics & Reported & Equiv. \\
\#23 & \href{https://github.com/PennyLaneAI/pennylane/issues/9510}{\#9510} & PennyLane & \code{decompose} pass on a \code{CSX} gate & Reported & Equiv. \\
\bottomrule
\end{tabular}
\end{table*}

\begin{figure}[t]
    \centering
    \begin{subfigure}{0.45\textwidth}
        \centering
        \begin{lstlisting}[style=coolqasm]
OPENQASM 2.0;
include "qelib1.inc";
gate cs q0,q1 {
p(pi/4) q0; cx q0,q1;
p(-pi/4) q1; cx q0,q1;
p(pi/4) q1; }
qreg q[5];
cs q[4],q[0];
        \end{lstlisting}
        \caption{Original QASM with \code{cs} gate definition.}
        \label{fig:bug_cs_original}
    \end{subfigure}
    \hfill
    \begin{subfigure}{0.48\textwidth}
        \centering
        \begin{lstlisting}[style=coolqasm]
OPENQASM 2.0;
include "qelib1.inc";
qreg q[5];
cs q[4], q[0];
# Error: 'cs' is not defined
# in this scope
        \end{lstlisting}
        \caption{BQSKit exported QASM with missing \code{cs} gate definition.}
        \label{fig:bug_cs_bqskit}
    \end{subfigure}
    \hspace{2mm}
    \caption{BQSKit omits the controlled phase gate (\code{cs}) definition when exporting to QASM (Bug~\IdBugMissingControlledPhaseCsGateDefinition{}).}
    \label{fig:bug_example_cs_gate}
\end{figure}

In addition to the bug illustrated in Listings~\ref{lst:bug_example_mcx_gate_py} and~\ref{lst:bug_example_mcx_gate_qasm}, which has been confirmed by the Pytket developers, we describe two more representative bugs discovered by \methodName{}.
Figure~\ref{fig:bug_example_cs_gate} demonstrates bug~\IdBugMissingControlledPhaseCsGateDefinition{} in BQSKit, where the platform omits the definition of a controlled phase gate (\code{cs}) when exporting to QASM. The figure contrasts the original QASM code containing the gate definition with BQSKit's exported version that lacks this crucial definition. When reimporting with BQSKit itself, this omission causes no issues, masking the bug in single-platform testing. However, when the QASM file is used with other platforms that require explicit gate definitions, the missing definition leads to compilation errors. The BQSKit developers acknowledged this as a complex issue that involves balancing standard gate definitions with compilation efficiency.

\begin{figure}[t]
    \centering
    \begin{subfigure}{0.45\textwidth}
        \centering
        \begin{lstlisting}[style=coolqasm]
OPENQASM 2.0;
include "qelib1.inc";
qreg q[11];
rxx(1.1761910836010856*pi) q[3],q[5];
        \end{lstlisting}
        \caption{Valid QASM generated by Pytket with \code{rxx} gate.}
        \label{fig:bug_rxx_pytket}
    \end{subfigure}
    \hfill
    \begin{subfigure}{0.48\textwidth}
        \centering
        \begin{lstlisting}[style=coolpython]
from qiskit.qasm2 import load
new_qc = load("valid_pytket_with_rxx.qasm")
# Error when importing with other platform:
# QASM2ParseError: "<input>:3,27: 'rxx' is not defined in this scope"
        \end{lstlisting}
        \caption{Crash when importing with Qiskit importer.}
        \label{fig:bug_rxx_qiskit}
    \end{subfigure}
    \hspace{2mm}
    \caption{Pytket generates valid QASM with \code{rxx} gate, but Qiskit's importer fails to recognize it (Bug~\IdBugMissingStandardRxxGateDefinition{}).}
    \label{fig:bug_example_rxx_gate}
\end{figure}

Figure~\ref{fig:bug_example_rxx_gate} shows bug~\IdBugMissingStandardRxxGateDefinition{}, where a valid QASM generated by Pytket includes an \code{rxx} gate.
Figure~\ref{fig:bug_rxx_qiskit} shows how Qiskit fails to recognize the \code{rxx} gate when importing the QASM code, raising a parse error. This bug represents a known limitation of Qiskit's importer, which becomes apparent through cross-platform testing. While the developers have provided a workaround using \code{qasm2.loads} with argument \code{custom\_instructions=LEGACY\_\-CUSTOM\_\-INSTRUCTIONS} rather than fixing the issue directly, this case demonstrates how \methodName{} effectively identifies gaps in platform interoperability.

Overall, \methodName{} finds both inconsistencies in the translation process from platform-specific representations to and from QASM and inequivalences between the resulting program pairs.
While the primary goal of \methodName{} is to detect bugs in the tested platforms, it also identified three bugs in the QCEC equivalence checking tool upon which our approach builds.
We reported these bugs to the QCEC developers, who confirmed the issues and provided fixes.

To avoid redundancy, we group multiple bugs pointing to the same root cause under a single GitHub issue.
We report bugs only if they have not been reported previously; only one bug found by \methodName{} had been reported before.
In a few cases (\TotalChangingConfiguration{}), after reporting a bug, we had to make configuration changes to continue with the evaluation without repeatedly triggering the same bug.
These changes included disabling specific gates, such as the \code{delay gate} (\IdBugUnrecognizedDelayGate{}) or \code{iswap} (\IdBugUnsupportedIswapGate{}), avoiding the use of classical registers (\IdBugDuplicateClassicalRegisterDefinition{}), or adding workarounds suggested by developers to prevent recurring bugs (\IdBugMissingStandardRxxGateDefinition{}).

\begin{answerbox}
  \textbf{Answer to RQ1}:
  \methodName{} effectively identifies bugs, uncovering \TotalAllBugs{} issues across the tested platforms, with \TotalConfirmedOrFixedBugs{} bugs already confirmed or fixed, demonstrating its practical impact.
\end{answerbox}

\subsection{RQ2: Impact of ITE Iterations}
\label{sec:rq_core_component}

We investigate how iteratively translating quantum circuits across different platforms affects the characteristics of the generated test programs and their effectiveness in detecting bugs.
One ITE iteration consists of creating an assembly program, importing it into a platform, applying a transformation, exporting to assembly, and adding the new assembly program to the equivalence class.
The number of ITE iterations is represented by $m$ in Algorithm~\ref{alg:qite_process}.

To quantify the impact of ITE iterations, we measure:
\begin{itemize}
\item \emph{Number of total gates per program:} This metric measures program complexity, as more complex programs are more likely to trigger subtle bugs.
\item \emph{Number of unique gates per program:} This metric captures the diversity of gates in the generated programs by counting the number of unique gate types.
For example, a \code{cx} gate is counted once, even if a program applies this kind of gate to different qubits.
\item \emph{Entropy of n-grams of instructions per program:} We compute the entropy of pairs and triplets of consecutive instructions in the generated programs to measure program diversity. Higher entropy indicates more diverse programs.
We consider instruction sequences as they appear in the assembly code since this is the input processed by the platforms.
\item \emph{Line coverage:} This metric assesses how much platform code is exercised by the generated programs.
\end{itemize}

We generate an initial set of 1,000 programs and perform five ITE iterations.
We repeat this experiment five times, consistent with prior work~\cite{xiaFuzz4AllUniversalFuzzing2024}, and then report the average and the confidence intervals.
Note that, for any experiment run, \methodName{} randomly selects a platform for each single program transformation, thus distributing the testing effort across the tested platforms in a balanced manner.

\begin{figure*}[t]
    \centering
    \begin{minipage}{0.49\textwidth}
        \centering
        \includegraphics[width=\linewidth]{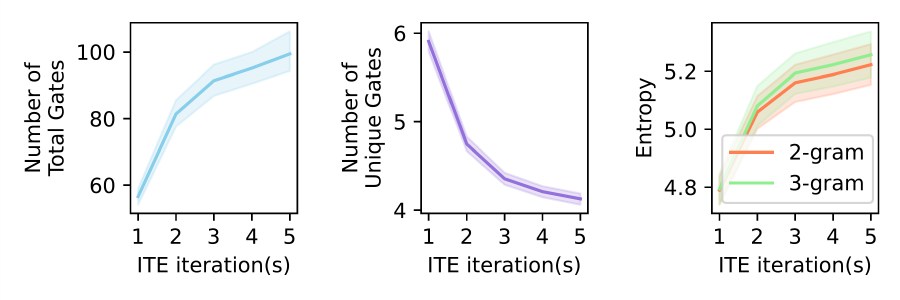}
        \caption{Program diversity versus number of ITE iterations, showing the average number of total gates (left), the number of unique gates (center), and the entropy of instruction n-grams (right) per generated program.}
        \label{fig:chain_length_vs_program_diversity}
    \end{minipage}
    \hfill
    \begin{minipage}{0.49\textwidth}
        \centering
        \includegraphics[width=\linewidth]{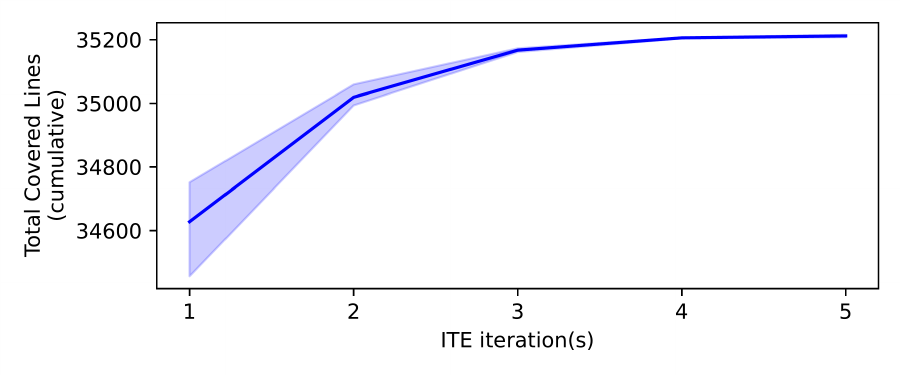}
        \caption{Cumulative line coverage versus number of ITE iterations.}
        \label{fig:chain_length_vs_coverage}
    \end{minipage}
\end{figure*}

Figure~\ref{fig:chain_length_vs_program_diversity} shows how program diversity changes with the number of ITE iterations.
Both the average number of total gates (left) and the entropy of instruction sequences (right) increase over iterations, with 3-grams showing slightly higher entropy levels but following similar trends.
This indicates that the \coreComponentName{} effectively generates more complex and diverse programs over time.
The average number of unique gates (middle), however, follows the opposite trend, decreasing over iterations.
This suggests that the platforms converge towards a common set of gates, which is expected due to the use of transformations that change the gate set to a small, universal gate set.
Indeed, the target gate set size of the transformations contains between two to five gates, depending on the platform, significantly reducing the gate set size average shown in Figure~\ref{fig:chain_length_vs_program_diversity}.

Figure~\ref{fig:chain_length_vs_coverage} shows the cumulative line coverage achieved on all platforms together across ITE iterations.
As the number of iterations increases, the cumulative line coverage also increases, indicating that the \coreComponentName{} continues to explore new code paths and potentially uncover additional bugs.

\begin{figure*}[t]
    \centering
    \begin{minipage}{0.43\textwidth}
        \centering
        \includegraphics[width=\linewidth]{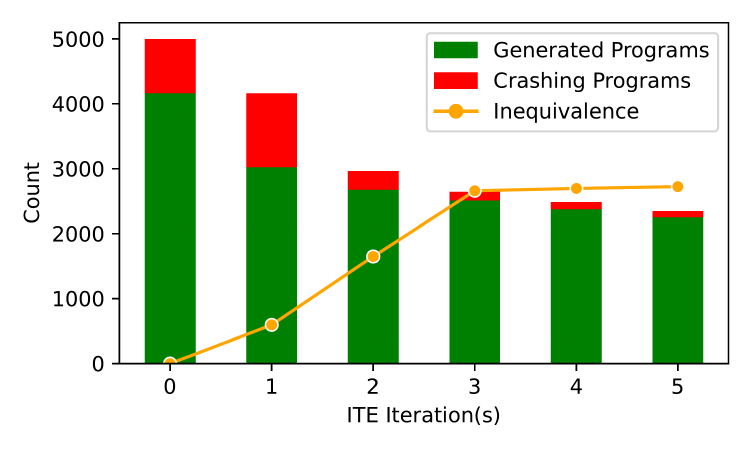}
        \caption{Number of programs and warnings detected over ITE iterations. The number of warnings is not cumulative and zero iterations correspond to crashes during generation, including the Python-to-QASM conversion.}
        \label{fig:chain_length_vs_programs_warnings}
    \end{minipage}
    \hfill
    \begin{minipage}{0.55\textwidth}
        \centering
        \includegraphics[width=\linewidth]{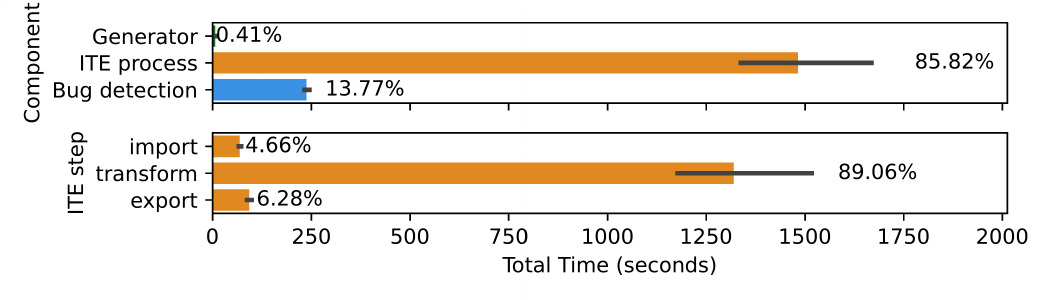}
        \caption{Time analysis of \methodName{} in terms of components (top) and steps of the \coreComponentName{} (bottom). Note the log scale.}
        \label{fig:time_analysis_components_and_functions}
    \end{minipage}
\end{figure*}

To better understand the impact of ITE iterations on program generation and bug detection, Figure~\ref{fig:chain_length_vs_programs_warnings} shows both the total number of programs and warnings detected over ITE iterations. At ITE iteration zero (initial generation), the process starts with direct program generation, including the Python-to-QASM conversion. While our target was to generate 5,000 programs (1,000 programs × 5 runs), the actual number is lower due to crashes during the conversion process - though these crashes themselves lead to the discovery of interesting bugs.
The vertical axis of Figure~\ref{fig:chain_length_vs_programs_warnings} shows (i) the number of generated programs per iteration, distinguishing between those successfully stored as assembly programs (green bar) and those that cause a crash (red bar); and (ii) the number of inconsistent pairs detected by the equivalence checker (yellow line).
Although \methodName{} usually runs the equivalence checker only at the end of the \coreComponentName{}, for this experiment we run it after each ITE iteration to understand its behavior over time. The number of detected inconsistencies initially grows but then stabilizes and slightly decreases in later ITE iterations.
This pattern emerges because the size of equivalence classes grows over time.

\begin{answerbox}
    \textbf{Answer to RQ2}:
    The iterations of the \coreComponentName{} are instrumental to effectively generate diverse test programs while increasing both program complexity and coverage.
\end{answerbox}

\subsection{RQ3: Efficiency of \methodName{}}

We study the efficiency of \methodName{} in terms of time spent in each component of \methodName{}, i.e., in the generator, the \coreComponentName{}, and bug detection, to generate and process a predefined number of programs.
We also measure the time spent in each of the steps of the \coreComponentName{}, where we consider the \code{import}, \code{transform}, \code{export} steps separately.
These metrics are relevant because they provide insights into the computational cost of each phase and the overall rate at which \methodName{} can generate test cases.
We use the same setup as used to study the \coreComponentName{} in Section~\ref{sec:rq_core_component}, consisting of generating 1,000 initial programs followed by $m=5$ ITE iterations.

Figure~\ref{fig:time_analysis_components_and_functions} shows the time analysis of \methodName{} components on top, and the steps of the \coreComponentName{} on the bottom.
The results indicate that the \coreComponentName{} consumes the most time, accounting for \PercentageTimeITEprocess{} of the total time of a typical run.
The bug detection component, which includes the equivalence checker, consumes \PercentageTimeBugDetection{} of the total time, whereas the generator component is the fastest, accounting for only \PercentageTimeGenerator{}.
This is understandable since the \coreComponentName{} component is the most complex and involves multiple steps and ITE iterations, including optimizations, while the generator component is based on a lightweight grammar-based approach.

In terms of steps of the \coreComponentName{}, the \code{transform} stage consumes the most time, accounting for \PercentageTimetransform{} of the total time, followed by the \code{export} stage (\PercentageTimeexport{}) and the \code{import} stage (\PercentageTimeimport{}).
This is expected since the \code{transform} stage often involves complex transformations, the \code{export} stage involves occasional conversion and serialization of gates that are not directly supported in QASM, and the \code{import} stage involves a more straightforward parsing of QASM code, thus being the fastest.

\begin{answerbox}
    \textbf{Answer to RQ3}:
    \methodName{} exhibits high efficiency, generating programs in just a fraction of a second (\AvgGeneratorTimeOverall{} seconds per program; generating \TotalProgramsGeneratorOverall{} programs over five runs in \TotalTimeGeneratorOverall{} seconds), allowing computational resources to be focused on the \coreComponentName{} and bug detection, which constitute \PercentageTimeITEprocess{} and \PercentageTimeBugDetection{} of the total time, respectively.
\end{answerbox}

\subsection{RQ4: Ablation Study}
\label{sec:ablation study}

\begin{figure*}[t]
    \centering
    \begin{subfigure}{0.48\textwidth}
        \includegraphics[width=\textwidth]{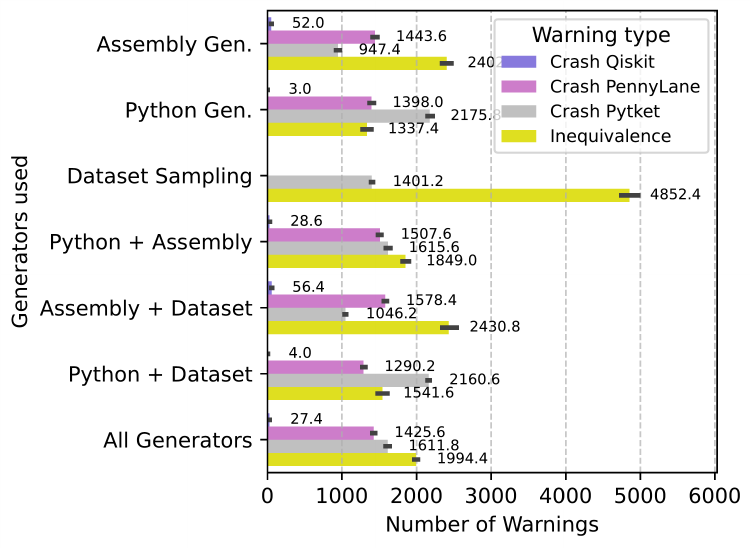}
        \caption{Bug detection by generator and platform.}
        \label{fig:ablation_bug_detection_generators}
    \end{subfigure}
    \hfill
    \begin{subfigure}{0.48\textwidth}
        \includegraphics[width=\textwidth]{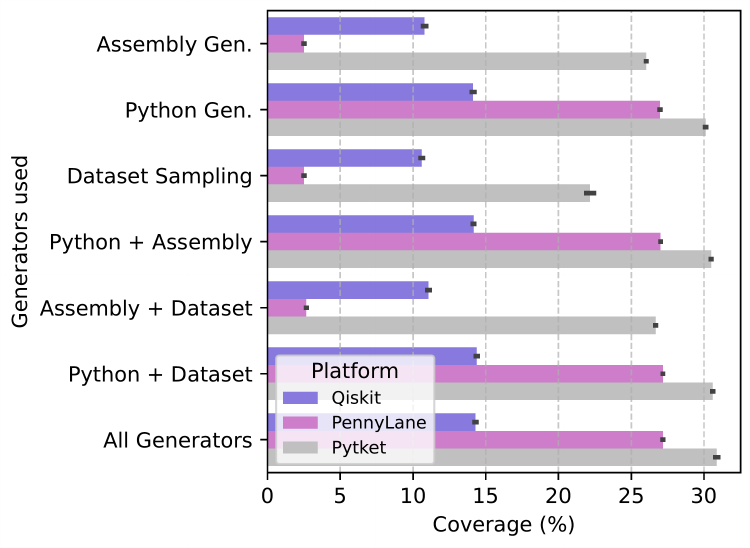}
        \caption{Coverage percentage by generator and platform.}
        \label{fig:ablation_study_coverage_generators}
    \end{subfigure}
    \caption{Ablation study results for different generators: (a) number of warnings detected and (b) coverage achieved across platforms.}
    \label{fig:ablation_study_generators}
\end{figure*}

\subsubsection{Impact of Program Generators}

First, we evaluate the impact of the three program generators (Section~\ref{sec:program_generation}) on the effectiveness of the approach.
We study seven setups: each generator individually, all pairs of generators, and all three generators combined.
For each setup, we generate 5,000 initial QASM programs and perform $m=5$ ITE iterations.
For a fair comparison, we configure all generators to produce programs with a maximum of 11 qubits and 15 gates.
Note that the ``sampling from dataset'' generator has a limited number of programs by design because it draws from a fixed dataset; once they are fully used, the remaining quota is filled by the other generators.
For example in our setup, the MQTBench dataset is fully used after about 500 programs that satisfy the size constraints mentioned above.
We measure the number of crash warnings, inconsistent pairs, and the code coverage achieved by each setup to assess their effectiveness.
We repeat the experiments five times and report the average and confidence intervals of our measurements.

Figure~\ref{fig:ablation_study_generators} shows the results of this ablation study.
In terms of triggering crashes, any combination involving the Python-based generator or the assembly-based generator is effective across all platforms, whereas the sampling from dataset strategy triggers crashes exclusively in Pytket. This might be due to the fact that the programs in MQTBench are known and could have been tested previously on Qiskit and PennyLane, thus reducing the likelihood of uncovering new crashes in those platforms.
In terms of inconsistencies, the sampling from dataset strategy produces the highest number of inconsistencies. Based on inspecting ten inconsistent pairs, eight are linked to PennyLane's exporter, which fails to preserve the target classical register of a measurement operation. The remaining two inconsistencies are related to Pytket reordering classical registers in the QASM code. We report this as a bug (\IdBugQasmRegisterReorderingCausesSemanticChanges{}), which the Pytket developers have confirmed.
Finally, in terms of code coverage, setups involving the Python-based generator achieve higher coverage because they exercise the converter code, covering more of the platform's code. The assembly-based and sampling from dataset strategy achieve similar coverage, with the assembly-based generator slightly outperforming the sampling from dataset strategy for Pytket.
Based on these results, we conclude that all three generators contribute to the effectiveness of \methodName{} in complementary ways, justifying their combined use in our approach.

\newcommand{\TransformImpactCrashIncrementInconsistentpair}{-466}
\newcommand{\TransformImpactCrashIncrementPennylane}{-522}
\newcommand{\TransformImpactCrashIncrementPytket}{724}
\newcommand{\TransformImpactCrashIncrementQiskit}{-85}
\newcommand{\TransformImpactCoverageIncrementPennylane}{0.91\%}
\newcommand{\TransformImpactCoverageIncrementPytket}{19.93\%}
\newcommand{\TransformImpactCoverageIncrementQiskit}{3.58\%}

\newcommand{\CoverageImprovementOnlychangeofgatestransforms}{4.4\%}
\newcommand{\CoverageImprovementOnlyoptimizertransforms}{42.8\%}
\newcommand{\CoverageImprovementBothtransforms}{43.7\%}

\subsubsection{Impact of Transformations}
Second, we evaluate the impact of the transformations applied by the \coreComponentName{} on the number of crashes, inconsistencies, and on the code coverage.
To this end, we compare \methodName{}, as described in Section~\ref{sec:approach}, with a version where all transformations are disabled.
We repeat the experiment five times and report the average percentage change.
In terms of code coverage, the transformations lead to an increase in all platforms: \TransformImpactCoverageIncrementPytket{} in Pytket, \TransformImpactCoverageIncrementPennylane{} in PennyLane, and \TransformImpactCoverageIncrementQiskit{} in Qiskit.
We also quantify the impact of optimizers and gate set changes on code coverage. Using only gate set changes yields a modest coverage increase (\CoverageImprovementOnlychangeofgatestransforms{}), while optimizers alone significantly boost coverage (\CoverageImprovementOnlyoptimizertransforms{}). Combining both achieves a \CoverageImprovementBothtransforms{} increase, showing their complementary contributions.

\newcommand{\EquivStrategyPrograms}{614}
\newcommand{\EquivStrategyFirstLastComparisons}{614}
\newcommand{\EquivStrategyRandomComparisons}{2,752}
\newcommand{\EquivStrategyFirstLastWarningPrograms}{367}
\newcommand{\EquivStrategyRandomWarningPrograms}{361}
\newcommand{\EquivStrategyBothWarningPrograms}{355}
\newcommand{\EquivStrategyNeitherWarningPrograms}{241}
\newcommand{\EquivStrategyFirstLastOnlyWarningPrograms}{12}
\newcommand{\EquivStrategyRandomOnlyWarningPrograms}{6}
\newcommand{\EquivStrategyFirstLastWarningCandidates}{367}
\newcommand{\EquivStrategyRandomWarningCandidates}{826}

\subsubsection{Impact of Equivalence-Comparison Strategy}
The equivalence oracle compares a subset of all pairs of programs within the same equivalence class to identify inconsistencies (Section~\ref{sec:bug_detection})
We compare two strategies for selecting these program pairs.
For this experiment, we run \methodName{} for five ITE iterations on 1,000 starting programs to obtain \EquivStrategyPrograms{} equivalence classes with at least two programs.
For each class, the approach constructs a transformation chain of the initial program and intermediate programs produced after each iteration.
The \emph{first-last} strategy compares only chain endpoints, while the \emph{random} strategy samples five pairs from each chain.
The first-last strategy performs fewer comparisons (\EquivStrategyFirstLastComparisons{} vs.\ \EquivStrategyRandomComparisons{}) and detects fewer inconsistencies (\EquivStrategyFirstLastWarningCandidates{} vs.\ \EquivStrategyRandomWarningCandidates{}), as it can miss faults involving intermediate programs.
We therefore use random pairing as the default in \methodName{}, and we keep first-last as a complementary configurable alternative.

\begin{answerbox}
    \textbf{Answer to RQ4}:
        All three program generators contribute complementarily to \methodName{}'s effectiveness.
        The transformations in the \coreComponentName{} are helpful for increasing coverage. For equivalence checking, random pair selection is a better tradeoff than checking only the first and last programs in each transformation chain because endpoint-only checking misses some warnings.
\end{answerbox}

\newcommand{\IterTestQProgramsPerHour}{10,162}
\newcommand{\QuteFuzzProgramsPerHour}{603}
\newcommand{\QDiffProgramsPerHour}{277}
\newcommand{\MorphQProgramsPerHour}{2,599}
\newcommand{\FuzzFourAllProgramsPerHour}{101}

\subsection{RQ5: Comparison with Prior Work}

We compare \methodName{} with prior approaches by running each approach for a total of 24 hours, using the same hardware setup for all approaches to ensure a fair comparison.
The comparison includes two kinds of baselines: single-platform approaches and multi-platform approaches.
\emph{Single-platform approaches} target one platform under test.
MorphQ~\cite{paltenghiMorphQMetamorphicTesting2023} tests Qiskit with metamorphic transformations and output-based oracles, so we run it for 24 hours on Qiskit.
Fuzz4All~\cite{xiaFuzz4AllUniversalFuzzing2024}, currently available only for Qiskit, uses an LLM-based program generator with two separate models: a stronger model for automatically refining generation prompts through autoprompting, and a cheaper model for the large-scale fuzzing loop. To match the computational budget of the other approaches, we run Fuzz4All for 24 hours using local models, specifically Gemma~4~8B for autoprompting and StarCoder2~3B for fuzzing. The same-platform differential testing oracle of Fuzz4All uses dynamic analysis to identify circuits created during program execution, and then checks QASM self-roundtrips and optimization-level consistency.

\emph{Multi-platform approaches} test multiple platforms, either sequentially or simultaneously.
QuteFuzz~\cite{iwumbweQuteFuzzFuzzingQuantum} generates quantum programs with features such as control flow, subcircuits, and measurements to fuzz quantum libraries, so we run it for eight hours each on Qiskit, Pytket, and Cirq.
QDiff~\cite{wangQDiffDifferentialTesting2021} mutates related circuits and compares execution-output distributions, so we update it for our environment and run it for 12 hours each on Qiskit and Cirq.
Because these baselines target Cirq, we added support for testing Cirq to \methodName{}.
\methodName{} runs once for 24 hours while testing Qiskit, PennyLane, Pytket, and Cirq simultaneously.
Note that the released baseline implementations were not designed for a Qiskit version as recent as the one used by \methodName{}, so we update their software stacks before this comparison to keep the platform versions aligned.

Table~\ref{tab:rq5_prior_work_comparison} summarizes the results.
We report the total number of warnings and the kind of oracle that has revealed them.
As inspecting all warnings is infeasible, we inspect a statistically relevant sample for each warning set, following prior bug-detection work that samples large warning populations and manually validates representative warnings~\cite{ghalebACheckerStaticallyDetecting2023}.
A single author annotates warnings until reaching the sample size required for a 90\% confidence level and a 10 percentage-point margin of error for each warning set, and clusters the sampled warnings by reduced program, warning message, and apparent root cause.
The table also records confirmed or known bugs when the warning classes can be mapped to developer-confirmed issues or existing bug reports.

\begin{table*}[t]
    \centering
    \footnotesize
    \caption{RQ5 comparison with prior quantum testing tools under a 24-hour budget. ``Other warnings'' denotes non-crash anomalies, including inequivalences for \methodName{} and statistical divergences for tools that use output-distribution oracles. The inspected warning samples use a 90\% confidence level and a 10 percentage-point margin of error.}
    \label{tab:rq5_prior_work_comparison}
    \setlength{\tabcolsep}{2pt}
    \begin{tabularx}{\textwidth}{@{}l>{\raggedright\arraybackslash}Xlrrrrrr@{}}
        \toprule
        \textbf{Tool} &
        \textbf{Platforms} &
        \textbf{Budget} &
        \textbf{Programs} &
        \multicolumn{2}{c}{\textbf{Warnings}} &
        \multicolumn{3}{c}{\textbf{Inspected Sample}} \\
        \cmidrule(lr){5-6}
        \cmidrule(lr){7-9}
        & & & & \textbf{Crash} & \textbf{Other} & \shortstack{\textbf{Sample}\\\textbf{Size}} & \shortstack{\textbf{Unique}\\\textbf{Warnings}} & \textbf{Bugs} \\
        \midrule
        \multicolumn{9}{l}{\textbf{Multi-platform}} \\
        \methodName{} &
        Qiskit, PennyLane, Pytket, Cirq &
        24h joint &
        243,886 &
        61,014 &
        46,050 &
        68 &
        33 &
        8 \\
        QuteFuzz~\cite{iwumbweQuteFuzzFuzzingQuantum} &
        Qiskit, Pytket, Cirq &
        8h/plat. &
        14,460 &
        84 &
        30 &
        43 &
        5 &
        3 \\
        QDiff~\cite{wangQDiffDifferentialTesting2021} &
        Qiskit, Cirq &
        12h/plat. &
        6637 &
        0 &
        0 &
        0 &
        0 &
        0 \\
        \midrule
        \multicolumn{9}{l}{\textbf{Single-platform}} \\
        MorphQ~\cite{paltenghiMorphQMetamorphicTesting2023} &
        Qiskit &
        24h &
        62,378 &
        11,967 &
        16 &
        68 &
        6 &
        2 \\
        Fuzz4All~\cite{xiaFuzz4AllUniversalFuzzing2024} &
        Qiskit &
        24h &
        2,413 &
        3 &
        0 &
        3 &
        1 &
        0 \\
        \bottomrule
    \end{tabularx}
\end{table*}

The fixed time budget also enables a direct comparative time analysis.
Normalizing by the total 24-hour budget, \methodName{} processes \IterTestQProgramsPerHour{} programs per hour, compared with \MorphQProgramsPerHour{} for MorphQ and \FuzzFourAllProgramsPerHour{} for Fuzz4All in the single-platform runs.
For the multi-platform baselines, the total rates across their platform-specific runs are \QuteFuzzProgramsPerHour{} programs per hour for QuteFuzz and \QDiffProgramsPerHour{} for QDiff.
These rates show that \methodName{} generates the largest number of tested programs per hour even while distributing its ITE workflow across four platforms.

The last column of Table~\ref{tab:rq5_prior_work_comparison} compares the approaches in terms of the number of unique bugs revealed while inspecting warnings.
It shows that \methodName{} identifies eight bugs, QuteFuzz three, MorphQ two, and neither QDiff nor Fuzz4All any in this comparison.
QuteFuzz and \methodName{} both test multiple platforms, but use different strategies.
QuteFuzz allocates its budget to fuzz each platform independently, while \methodName{} uses the same 24-hour budget to repeatedly pass programs across platform importers, transformations, and exporters to check equivalence classes.
The bugs they expose in Cirq reveal this complementarity:
QuteFuzz discovers an optimization bug where gates before measurements are folded into that measurement, causing later measurements of the same qubit to observe incorrect state.
\methodName{} discovers a different interoperability bug: Qiskit emits lowercase \code{u} in QASM gate definitions, which Cirq's importer rejects.
MorphQ identifies two pre-existing Qiskit bugs.
We conjecture QDiff's lack of success in our setup stems from simplistic mutations, such as canceling gates already handled by recent platform versions.
For Fuzz4All, using local models matched to our GPU capacity rather than frontier models may explain its lack of success, as larger LLMs are more effective.

\begin{answerbox}
    \textbf{Answer to RQ5}:
        Under the same 24-hour budget, \methodName{} complements prior single- and multi-platform quantum testing tools by jointly exercising Qiskit, PennyLane, Pytket, and Cirq through ITE, processing the most programs per hour, and producing the largest number of unique warnings in the inspected samples.
\end{answerbox}

\section{Threats to Validity}
We identify five main threats to the validity of our work.
First, \methodName{}'s applicability to new quantum computing platforms relies on their support for importing and exporting assembly code, particularly QASM, one of several quantum program representations.
Although this is common in major platforms, not all quantum computing platforms support it, limiting \methodName{}'s generalizability.
Cirq illustrates a related compatibility threat: even when a platform supports QASM import and export, small dialect differences may require local normalization, such as preserving unused qubits or translating \code{u} to \code{u3}.
Second, \methodName{} depends on semantics-preserving transformations provided by the platforms.
During our evaluation, we discovered that some platform-provided transformations documented as semantics-preserving actually change program semantics, so we manually check each transformation type.
Third, if a platform is significantly buggier than others, it can affect our evaluation by prematurely terminating the ITE loop.
For example, BQSKit's reliability issues led us to exclude it from RQ2--RQ5, as many transformation attempts failed due to platform-specific bugs.
This suggests that \methodName{} is most effective when testing platforms of comparable reliability.
Fourth, our implementation and experimental scripts may contain bugs that affect the reported results.
We mitigate this risk through manual inspection of reduced failures and repeated experiments that report averages and confidence intervals.
Fifth, our generated and sampled benchmark programs may not represent all quantum programs used in practice; we mitigate this risk by combining three different program generators, including one that samples from a real-world dataset.

\section{Related Work}

\paragraph{Quantum Testing and Debugging}
Quantum software engineering~\cite{murilloQuantumSoftwareEngineering2025,zhaoQuantumSoftwareEngineering2021} and quantum software testing and analysis~\cite{paltenghiSurveyTestingAnalysis2024} are rapidly growing fields.
Empirical studies highlight the prevalence of quantum-specific bugs in quantum software~\cite{paltenghiBugsQuantumComputing2022, luoComprehensiveStudyBug2022a}.
To address these issues, researchers have proposed static analyses~\cite{paltenghiAnalyzingQuantumPrograms2024,kaulUniformRepresentationClassical2023,zhaoQCheckerDetectingBugs2023}, dynamic analyses~\cite{10172808}, delta debugging~\cite{pontolilloDeltaDebuggingPropertyBased2024}, mutation testing~\cite{mendiluzeMuskitMutationAnalysis2022}, and metamorphic testing~\cite{abreuMetamorphicTestingOracle2022}.
Other approaches leverage runtime information from executions on quantum computers, using runtime projections and statistical assertions~\cite{liProjectionbasedRuntimeAssertions2020,liuQuantumCircuitsDynamic2020,huangStatisticalAssertionsValidating2019, kangStatisticalTestingQuantum2024}.
Debugging quantum programs is particularly challenging~\cite{miranskyyTestingDebuggingQuantum2021}, mainly due to the difficulty of measuring quantum states without disrupting computation.
To address these challenges, researchers have proposed runtime assertions~\cite{liProjectionbasedRuntimeAssertions2020,liuQuantumCircuitsDynamic2020}, statistical assertions~\cite{huangStatisticalAssertionsValidating2019, kangStatisticalTestingQuantum2024}, non-destructive discrimination methods~\cite{liuSystematicApproachesPrecise2021}, and interactive debugging tools~\cite{metwalliCirquoSuiteTesting2023}.
While these approaches focus on quantum applications, \methodName{} targets quantum computing platforms.

\paragraph{Testing Quantum Computing Platforms}
Several fuzzers address the challenges of testing quantum computing platforms.
QDiff~\cite{wangQDiffDifferentialTesting2021} performs differential testing~\cite{mckeemanDifferentialTestingSoftware1998} across optimization levels and backend configurations within a single platform, or across simulators~\cite{blackwellFuzzingBasedDifferentialTesting2024}.
MorphQ~\cite{paltenghiMorphQMetamorphicTesting2023} uses metamorphic testing~\cite{chenMetamorphicTestingNew2020} to compare outputs across backend configurations in Qiskit.
Fuzz4All~\cite{xiaFuzz4AllUniversalFuzzing2024} also targets Qiskit, employing a generator based on large language models and code examples.
Unlike these single-platform approaches, \methodName{} is a multi-platform testing technique for quantum software.
UPBEAT~\cite{huUPBEATTestInput2024} targets language-level input validation bugs in Q\# libraries.
Our core contribution, the novel \coreComponentName{}, can be combined with existing program generators, including those of MorphQ~\cite{paltenghiMorphQMetamorphicTesting2023} and Fuzz4All~\cite{xiaFuzz4AllUniversalFuzzing2024}.
FuzzQ~\cite{klimis2025shaking} and work by Blackwell et al.~\cite{blackwellFuzzingBasedDifferentialTesting2024} perform cross-platform differential testing, but unlike \methodName{}, they compare simulator behavior instead of testing cross-platform interoperability.
A difference to all the above is that \methodName{} uses cross-platform interoperability as a means to generate new quantum programs.

\paragraph{Quantum Program Equivalence}
Generating equivalent quantum programs is central to our equivalence oracle.
Prior work has explored this, often at a smaller scale~\cite{xuQuartzSuperoptimizationQuantum2022}.
\citet{moriQuantumCircuitUnoptimization2024} propose unoptimization techniques to benchmark optimization tools.
QDiff~\cite{wangQDiffDifferentialTesting2021} applies gate equivalence transformations to create equivalent programs for cross-backend comparison within a single platform.
While these approaches focus on single-platform equivalence, their transformations could be integrated into our \coreComponentName{}.
MorphQ~\cite{paltenghiMorphQMetamorphicTesting2023} uses metamorphic transformations to generate pairs of semantically equivalent programs, but not all are semantics-preserving, making them unsuitable for direct use in our approach.
Other methods for creating equivalent circuits include circuit mapping techniques~\cite{shafaeiQubitPlacementMinimize2014, cowtanQubitRoutingProblem2019a, zulehnerEfficientMappingQuantum2018, willeMappingQuantumCircuits2019, zhangTimeoptimalQubitMapping2021, tannuNotAllQubits2019}, which could also be incorporated to transform circuits.
Unlike prior work that relies on external code, our approach focuses on built-in platform transformations, making them part of the code under test.
Equivalence checking is essential for verifying that program transformations preserve semantics, including QCEC~\cite{pehamEquivalenceCheckingQuantum2022}, which is based on the ZX-calculus~\cite{coeckeInteractingQuantumObservables2011, kissingerPyZXLargeScale2020}, and the automata-based AutoQ~\cite{chenAutomataBasedFrameworkVerification2023, chenAutoQAutomataBasedQuantum2023}.
Giallar~\cite{tao2022giallar} verifies that compiler passes preserve the semantics of circuits, but unlike \methodName{}, focuses on a single optimization pass at a time.

\paragraph{Quantum Representation and Interoperability}
A common intermediate representation (IR) is essential for cross-platform testing.
While emerging IRs such as OpenQASM3~\cite{crossOpenQASMBroaderDeeper2022}, Quil~\cite{smithPracticalQuantumInstruction2017}, QIR~\cite{QirallianceQirspec2025}, and QASM-HL~\cite{javadiabhariScaffCCFrameworkCompilation2014} exist, QASM remains the most widely adopted.
Our approach leverages QASM2 for iterative import and export, combined with platform-specific transformations.
The value of a common IR has also been demonstrated in deep learning testing~\cite{liuNNSmithGeneratingDiverse2023}, though it is typically used for one-way translation, unlike our iterative \coreComponentName{}.
Even with standardized IRs, interoperability challenges persist.
For example, a recent study reports that 75\% of defects in deep learning occur during operator conversion, with 33\% leading to semantically inequivalent models~\cite{jajalInteroperabilityDeepLearning2024}.
Our work addresses similar challenges for quantum computing and QASM.

\paragraph{Compiler and Runtime Engine Testing}
Our work relates to compiler testing~\cite{chenSurveyCompilerTesting2020}, as quantum computing platforms conceptually resemble traditional compilers.
Grammar-based generators have been used to test compilers and similar systems, including the Java Virtual Machine~\cite{sirerUsingProductionGrammars2000}, C compilers~\cite{yangFindingUnderstandingBugs2011, even-mendozaGrayCGreyboxFuzzing2023}, SMT solvers~\cite{wintererValidatingSMTSolvers2024}, database engines~\cite{riggerTestingDatabaseEngines2020}, and deep learning infrastructure~\cite{maFuzzingDeepLearning2023, liuNNSmithGeneratingDiverse2023, wangDeepLearningLibrary2020, luoGraphbasedFuzzTesting2021}.
However, prior compiler testing techniques do not address the unique challenges of cross-platform testing in quantum computing.
Equivalence-modulo-input (EMI)~\cite{leCompilerValidationEquivalence2014} and work on testing the MLIR compiler infrastructure~\cite{liangfinding} shares the idea of testing transformation systems by producing equivalent programs. \methodName{} differs by deriving the equivalent programs via cross-platform interoperability.
Unlike EMI, which produces programs that behave equivalently on a specific input, \methodName{} generates programs that are semantically equivalent across all inputs.
This design is more suitable for testing quantum computing platforms where checking for input-specific equivalence is challenging due to the non-determinism of executing quantum programs.

\section{Conclusion}
We introduce \methodName{}, a novel cross-platform testing technique for quantum platforms.
\methodName{} uses quantum assembly code for program generation and equivalence checking, enabling cross-platform comparisons.
Evaluation on five platforms revealed \TotalAllBugs{} bugs, with \TotalConfirmedOrFixedBugs{} confirmed or fixed.
The core component, our novel \coreComponentName{}, generates diverse programs while testing multiple platforms simultaneously.
Given the growing importance and diversity of quantum software platforms, \methodName{} provides an effective and efficient approach for quantum platform testing.

\section{Data Availability}
\label{sec:data_availability}

We make all data, source code, and experimental results publicly available at \url{https://github.com/sola-st/IterTestQ-quantum-platform-testing/} and \url{https://doi.org/10.6084/m9.figshare.33016415}.

\section{Acknowledgments}
This work was supported by the German Research Foundation (DFG; projects 492507603, 516334526, and 526259073).

\bibliographystyle{ACM-Reference-Format}
\bibliography{phd-mattepalte,moreRefs}

\end{document}